\documentclass[11pt]{article}
\pdfoutput=1

\usepackage{euscript}
\usepackage{amssymb}
\usepackage{amsfonts}
\usepackage{amsbsy}
\usepackage{epsfig}
\usepackage{amsthm}
\usepackage{amscd}
\usepackage{amstext}
\usepackage{verbatim}
\usepackage{amsmath}
\usepackage{cancel}
\usepackage{capt-of}
\usepackage{empheq}

\usepackage{color}

\def\ben{\begin{equation}}
\def\een{\end{equation}}
\def\half{{\textstyle{1\over2}}}

\let\a=\alpha \let\b=\beta \let\g=\gamma

\let\s=\sigma

\def\be{\begin{equation}}
\def\ee{\end{equation}}
\def\beq{\begin{equation}}
\def\eeq{\end{equation}}
\def\ba{\begin{array}}
\def\ea{\end{array}}

\def\dalemb#1#2{{\vbox{\hrule height .#2pt
       \hbox{\vrule width.#2pt height#1pt \kern#1pt
               \vrule width.#2pt}
       \hrule height.#2pt}}}

\newcommand{\bea}{\begin{eqnarray}}
\newcommand{\eea}{\end{eqnarray}}
\newcommand{\tr}{{\rm tr} }
\newcommand{\Tr}{{\rm Tr} }

\def\Z{{{\Bbb Z}}}

\begin{document}

\begin{center}

{ \Large {\bf
Large $N$ matrices from a nonlocal spin system
}}

\vspace{1cm}

Dionysios Anninos${}^\sharp$, Sean A. Hartnoll${}^\flat$, Liza Huijse${}^\flat$ and Victoria L. Martin${}^\flat$

\vspace{1cm}

{\small
{\it ${}^\sharp$ School of Natural Sciences, Institute for Advanced Study, \\
Princeton, NJ 08540, USA. \\
${}^\flat$ Department of Physics, Stanford University, \\
Stanford, CA 94305-4060, USA }}

\vspace{1.6cm}

\end{center}

\begin{abstract}

Large $N$ matrices underpin the best understood models of emergent spacetime.
We suggest that large $N$ matrices can themselves be emergent from simple quantum mechanical
spin models with finite dimensional Hilbert spaces. We exhibit the emergence of large $N$ matrices
in a nonlocal statistical physics model of order $N^2$ Ising spins. The spin partition function is shown to admit
a large $N$ saddle described by a matrix integral, which we solve. The matrix saddle is dominant at high temperatures,
metastable at intermediate temperatures and ceases to exist below a critical order one temperature.
The matrix saddle is disordered in a sense we make precise and competes with ordered low energy states.
We verify our analytic results by Monte Carlo simulation of the spin system.

\end{abstract}

\pagebreak
\setcounter{page}{1}

\section{Introduction}

Large $N$ by $N$ matrices \cite{'tHooft:1973jz} are ubiquitous in contemporary fundamental physics. In particular, the currently best understood models of emergent spacetime involve `microscopic' degrees of freedom that are valued in large $N$ matrices \cite{Klebanov:1991qa, Ginsparg:1993is, Banks:1996vh, Maldacena:1997re}. 

While the large $N$ matrices may be taken to be the microscopic starting point for quantum gravity, in this paper we explore the possibility that the large $N$ matrices are themselves emergent from an even simpler model, with a finite dimensional space of states. Our eventual hope is that if spacetime can emerge from, for instance, a quantum mechanical spin system, then it may be possible to understand phenomena such as emergent locality in detail using the many quantum information-theoretic results known for such systems.

Our work can also be motivated from a statistical or many-body physics perspective. Building on the classic paper \cite{Brezin:1977sv}, a large body of techniques exists for solving matrix integrals and matrix quantum mechanics. If certain spin systems can be mapped onto large $N$ matrices, then they may be exactly soluble. Furthermore, novel types of phase transitions may occur that do not involve symmetry breaking but rather changes in topological order (cf. \cite{Wen:2004ym}) such as the connectivity of the eigenvalue distribution of the matrices \cite{Brezin:1977sv}.

The simplest class of matrix theories are matrix integrals. These are statistical physics partition functions with no time. Our model will be of this type. An important precedent for our work is \cite{parisi}, who relate the high temperature phase of a nonlocal spin system to a matrix integral and furthermore show that the low temperature phase is glassy. The model we will solve is in fact the `ferromagnetic' version of the `antiferromagnetic' model studied in \cite{parisi}. Some of our manipulations will parallel \cite{parisi} closely while others will be different. We will see that the ferromagnetic model has its own charm.

Our starting point is the nonlocal Ising spin partition function (\ref{eq:original}) for $N \times N_F \sim N^2$ spins. We show that as $N \to \infty$ the partition function can be written as the matrix integral (\ref{eq:Zfinal}), with a parameter determined self-consistently by the constraint (\ref{eq:constraint}). The emergence of an $SO(N)$ symmetry is nontrivial, as is the fact that the matrix action is purely single trace. We solve the matrix integral by finding the eigenvalue distribution (\ref{eq:AB}). We verify our results via Monte Carlo simulation in figure \ref{fig:MC}. The matrix integral saddle is found to be metastable below a temperature $T_* \sim N$ and ceases to exist below a critical temperature $T_c \sim 1$. A certain correlation between sets of four spins (\ref{eq:niceX}) is shown to distinguish the disordered matrix integral saddle from competing ordered low energy states.
We discuss possible implications of these results and future directions in the final section.

\section{A nonlocal Ising model}

Let us introduce the model. The basic operators in the theory are an $N\times N_F$ matrix worth of spins $\vec S^{Ab}$. Here the matrix indices $A = 1 \ldots N$ and $b = 1 \ldots N_F$. Without loss of generality we can take $N_F \leq N$, but in general we will imagine that $N_F \sim N$ are the same order of magnitude, so that
\be
\alpha \equiv \frac{N_F}{N} \,,
\ee
is order one. We will have in mind spin-half representations, normalized so that each $S_i^{Ab}$ has eigenvalues $\sigma^{Ab} = \pm 1$. With a view to connecting with bosonic matrices later (whose effective action will have an emergent $SO(N)$ symmetry), we want to write down Hamiltonians with a trace structure. The simplest nontrivial model is the nonlocal `matrix Ising model'
\be\label{eq:isingH}
H_\text{Ising} = \lambda \, \tr \left(S_z S^T_z S_z S_z^T \right) = \lambda \, S^{Ab}_z S^{Cb}_z S^{Cd}_z S^{Ad}_z  \,.
\ee
When $\lambda < 0$ the model is `ferromagnetic'; the four-spin interaction favors an even number of spins to be aligned and, in particular, the states with all spins aligned (but not only these states, as we discuss below) minimize the energy. The `antiferromagnetic' $\lambda > 0$ case, however, favors an odd number of each four-spins to be aligned.
In this paper we study the ferromagnetic case with $\lambda < 0$.
The antiferromagnetic model exhibits frustration and glassiness and was studied in \cite{parisi}.

The Hamiltonian (\ref{eq:isingH}) has several symmetries. The energy is invariant under permuting rows or columns, this gives an $S_N \times S_{N_F}$ symmetry. Furthermore, the Hamiltonian is invariant under flipping the sign of all the spins in a single row or column, this gives a $\Z_2^N \times \Z_2^{N_F}$ symmetry. These symmetries do not determine the Hamiltonian to be of the form (\ref{eq:isingH}), even when restricted to four-spin interactions. For instance, these symmetries allow the matrix indices to be repeated more than twice. In this sense the trace structure of (\ref{eq:isingH}) is fine tuned. The `matrix' structure in (\ref{eq:isingH})
does not imply $SO(N)$ or $SO(N_F)$ symmetries, which will be emergent in certain large $N$ phases.

The Hamiltonian (\ref{eq:isingH}) is reminiscent of that for a $\Z_2$ gauge theory. It differs in the nonlocality of the four-spin interactions. In particular, the Hamiltonian does not have the full local $\Z_2^{N N_F}$ symmetry of a gauge theory. Nonetheless, the $\Z_2^{N+N_F}$ symmetry we have described is still much larger than an overall global $\Z_2$ symmetry.

The quantity of concern to us will be the ferromagnetic, finite temperature partition function. This is
\be\label{eq:original}
Z(\beta) = \Tr \, e^{- \beta H} = \sum_{\sigma \in \{\pm 1\}^{N N_F}} e^{\frac{\beta}{16 N} \, \s^{Ab} \s^{Cb} \s^{Cd} \s^{Ad}} \,.
\ee
We have chosen a convenient normalization, $\lambda = - 1/(16N)$. Here we use $\Tr$ to denote the trace over the spin Hilbert space (rather than the matrix index trace $\tr$).

\subsection{Hubbard-Stratonovich transformations}

The sum over spins can be exactly reformulated as an integral over bosonic degrees of freedom. This is achieved in two steps.
Introducing a symmetric $N \times N$ matrix $Q_{AB}$ as a Hubbard-Stratonovich field, we can write
\bea
Z(\beta) & = & \frac{1}{2^{N/2}} \left(\frac{2 N}{\pi\beta}\right)^{N(N+1)/4}  \sum_{\sigma \in \{\pm 1\}^{N N_F}} \int dQ e^{- \frac{N}{\beta} Q_{AB} Q_{AB} + \frac{1}{2} \,  Q_{AB} \s^{Ac} \s^{Bc}} \\
& = & \frac{1}{2^{N/2}} \left(\frac{2 N}{\pi\beta}\right)^{N(N+1)/4} \int dQ e^{- \frac{N}{\beta} Q_{AB} Q_{AB}} \left(\sum_{\sigma \in \{\pm 1\}^{N}} e^{ \frac{1}{2}  \, Q_{AB} \s^{A} \s^{B}} \right)^{N_F} \label{eq:second} \\
& = &\frac{1}{2^{N/2}} \left(\frac{2 N}{\pi\beta}\right)^{N(N+1)/4}\int dQ e^{- \frac{N}{\beta} Q_{AB} Q_{AB} + N_F \log z(Q)} \,. \label{eq:third}
\eea
For the normalization of the first line, note that the integral is over symmetric matrices while the sum over $A$ and $B$ in the exponent runs over all indices. In a mean field limit one would have $Q_{AB} = \frac{\beta}{4 N} \s^{Ac} \s^{Bc}$, we can roughly have this in our mind as the physical meaning of the matrix $Q$.
The second line uses a step adapted from the solution of the Sherrington-Kirkpatrick model of spin glasses, see e.g. \cite{Denef:2011ee}. In the third line we introduced the reduced partition function for $N$ (rather than $N \times N_F$) Ising spins
\be\label{eq:reduced}
z(Q) = \sum_{\sigma \in \{\pm 1\}^{N}} e^{ \frac{1}{2}  \, Q_{AB} \s^{A} \s^{B}} \,.
\ee

To get rid of the sum over spin states entirely, we perform a second Hubbard-Stratonovich transformation. This is also a standard manipulation, e.g. \cite{Denef:2011ee}. The reduced partition function (\ref{eq:reduced}) can be written as
\bea
z(Q) & = & \frac{1}{(2\pi)^{N/2}} \sum_{\sigma \in \{\pm 1\}^{N}} \frac{1}{\sqrt{\det Q}} \int d w \, e^{- \frac{1}{2} w^T \cdot Q^{-1} \cdot w - w^B \sigma^B} \\
& = & \left(2/\pi\right)^{N/2} \frac{1}{\sqrt{\det Q}} \int d w \, e^{- \frac{1}{2} w^T \cdot Q^{-1} \cdot w + \sum_B \log\cosh \left(w^B \right)} \,.\label{eq:logcosh}
\eea
Here we have introduced an $N$-component bosonic mode $w^A$. The above expression is exact. Inserting this expression into the full partition function (\ref{eq:third}), we have recast the spin system as one matrix degree of freedom $Q$ interacting with $N_F$ vector degrees of freedom $w$. However, while these bosonic degrees of freedom are matrices and vectors under an $SO(N)$ action, the final term in the exponent in (\ref{eq:logcosh}) is not $SO(N)$ invariant. The objective in the next few steps is to deal with this term.

Taylor expanding the non-$SO(N)$ invariant term in (\ref{eq:logcosh}) gives
\be\label{eq:zqexpand}
z(Q) = \left(2/\pi\right)^{N/2} \frac{1}{\sqrt{\det Q}} \int d w \, e^{- \frac{1}{2} w^T \cdot Q^{-1} \cdot w + \sum_B \left( \frac{1}{2} (w^B)^2 - \frac{1}{12} (w^B)^4 + \cdots \right)} \,.
\ee
We would like to be able to keep only the first (quadratic in $w$) term in the expansion of $\log \cosh$ above, by arguing that the higher order non-singlet terms are subleading in a large $N$ expansion. Being able to neglect the non-singlet terms is the most nontrivial step in relating the spin system to a matrix integral. Further manipulations are necessary before we can do this, as we explain shortly. If we were to simply go ahead and neglect the quartic and higher terms in (\ref{eq:zqexpand}), then the remaining quadratic integral over $w$ would be easily performed to give
\be\label{eq:naive}
z(Q) \stackrel{N\to\infty}{=} 2^N \frac{1}{\sqrt{\det Q}}\frac{1}{\sqrt{\det\left(Q^{-1} - 1 \right)}} = \frac{2^N}{\sqrt{\det\left(1 - Q \right)}}= 2^N e^{- \frac{1}{2} \tr \log \left( 1 - Q \right)} \,,
\ee
This expression beautifully re-expresses the sum over spins (\ref{eq:reduced}) as a single trace exponent. However, it is not quite correct.

To see the problem with (\ref{eq:naive}), consider a formal high temperature (i.e. $\beta \to 0$) expansion of the full partition function to all orders. From the partition function (\ref{eq:third}) we expect that inside the integral $Q_{AB} \sim \sqrt{\beta/N}$ at large temperatures, as the quadratic term in the action becomes dominant. This holds for all components of $Q$, diagonal and off-diagonal. However, the propagator of the $w$ modes in (\ref{eq:zqexpand}), read off from the quadratic $\frac{1}{2} w^T \cdot ( Q^{-1} - 1) \cdot w$ term, is
\be\label{eq:G}
G = \frac{Q}{1 - Q} \,.
\ee
The diagonal and off-diagonal components of this matrix scale differently, and we shall see that this causes problems. In particular, order by order in $\beta$, that is order by order in an expansion of $G$ in powers of $Q$, the $N$ scaling takes the form $G_{AB} \sim s \, \delta_{AB} + t/\sqrt{N}$, where the coefficients $s$ and $t$ are independent of $N$. Thus while the off-diagonal components of $G_{AB}$ scale like $1/\sqrt{N}$, the diagonal components are order one. These order one diagonal components follow immediately from the Gaussian expectation value 
$\langle (Q^{2n})_{AB} \rangle \propto \delta_{AB}$. The coefficient of proportionality is order one because the normalization of the quadratic term in (\ref{eq:third}) implies that $\langle\tr [Q^{2n}]\rangle \sim N$.

The $SO(N)$ invariant terms in (\ref{eq:zqexpand}) resum into an exponent of order $N$ in (\ref{eq:naive}). How big is the contribution from the neglected non-singlet terms? Expanding out the non-singlet terms in (\ref{eq:zqexpand}) and Wick contracting with the $w$ propagator (\ref{eq:G}) gives terms such as $G_{AA}^2, G_{AB}^4, G_{AB}^2 G_{AC}^2 G_{BC}^2$, etc. We can ask how a term with $2 m$ factors of $G$ will scale with $N$. Let us suppose all the components of $G$ were to scale like $1/\sqrt{N}$ (which is not the case yet). Noting that the number of sums over indices is at most $m$ (because the terms in (\ref{eq:zqexpand}) are $w_A^4, w_A^8$, etc. and so each index is repeated four times or more for every two factors of $G$), we conclude that the terms scale at most as $N^{-(2m)/2 + m} \sim 1$. These terms will then exponentiate into an order one exponent that is subleading compared to the order $N$ exponent in (\ref{eq:naive}). This is what we would like to happen.

The diagonal components of $G$, however, are not suppressed at large $N$. This leads to terms such as $G_{AA}^2 \sim N$ that compete with the singlet terms kept in (\ref{eq:naive}). To fix (\ref{eq:naive}) we therefore need to remove the order one trace from $G$. This can be done as follows \cite{parisi}. From the definition of the reduced partition function (\ref{eq:reduced}), we can write
\be
z(Q) = e^{\frac{N(\gamma-1)}{2}} z(\widetilde Q) \,,
\ee
where we removed some part of the trace of $Q$, so that
\be
\widetilde Q_{AB} = Q_{AB} - (\gamma - 1) \delta_{AB} \,,
\ee
for any constant $\gamma$. The notation $(\gamma - 1)$ in the above two formulae is for future convenience. In particular, if we neglect the non-singlet terms as in (\ref{eq:zqexpand}), but now for $z(\widetilde Q)$ rather than $z(Q)$, we obtain the (different to (\ref{eq:naive})) new answer
\be\label{eq:correct}
z(Q) \stackrel{N\to\infty}{=} 2^N e^{\frac{N(\gamma-1)}{2}} e^{- \frac{1}{2} \tr \log \left( \gamma - Q \right)} \,. 
\ee
It is now clear what we have to do, we must choose $\gamma$ so that the propagator $\widetilde G$ for $\widetilde w$ (the modes that decouple the partition function $z(\widetilde Q)$) satisfies
\be\label{eq:tracecondition}
\tr \, \widetilde G = \tr \frac{\widetilde Q}{1 - \widetilde Q} = - N + \tr \frac{1}{\gamma - Q} = 0 \,.
\ee
By removing the trace of $\widetilde G$ in this way, all terms in $\widetilde G$ will scale like $1/\sqrt{N}$ and therefore (\ref{eq:correct}) will be true, following the $N$ counting argument of the previous paragraph.\footnote{The order one diagonal elements now introduced into $\widetilde Q$ do not disrupt this argument. These terms do not change the $1/\sqrt{N}$ scaling of $\widetilde G$.}
Having obtained (\ref{eq:correct}), the full large $N$ partition function becomes
\be\label{eq:Zfinal}
Z(\beta) \stackrel{N\to\infty}{=} \left(\frac{2 N}{\pi\beta}\right)^{N^2/4} e^{\alpha N^2 \left[ \frac{\gamma-1}{2} + \log 2 \right]}
\int dQ e^{- N \tr \left[ \frac{1}{\beta} Q^2 + \frac{\alpha}{2} \log \left(\gamma -   Q \right) \right]} \,.
\ee
Here we have only kept the leading order in large $N$ terms in the prefactor. With the large $N$ partition function in this form, we can now apply the highly developed machinery of matrix integrals to our statistical physics problem \cite{Brezin:1977sv}. The large $N$ matrices have emerged from the nonlocal spin system. The partition function has acquired an emergent $SO(N)$ invariance and is furthermore described by a single trace action.

At this point we can note (following \cite{parisi}) that the value of $\gamma$ that we have chosen in (\ref{eq:tracecondition}), is precisely the one that extremizes the exponent in the matrix integral (\ref{eq:Zfinal}). Namely
\be\label{eq:constraint}
\left\langle \tr \frac{1}{\gamma - Q} \right\rangle = N \,.
\ee
Here $\langle \rangle$ refers to the matrix integral average. Thus (\ref{eq:constraint}) has the form of a self-consistent `gap equation' for $\gamma$. We feel that the physical interpretation of this constraint remains to be fully elucidated.

In the following section we proceed to solve the matrix integral (\ref{eq:Zfinal}), with $\gamma$ self-consistently determined by the constraint (\ref{eq:constraint}). We note in passing that a matrix integral similar to (\ref{eq:Zfinal}) arises as the `radial subsector' of certain multi-matrix integrals \cite{Masuku:2014wxa}.

\subsection{Matrix model saddle}

Rotating to an eigenvalue basis, the partition function (\ref{eq:Zfinal}) becomes
\be\label{eq:Zx}
Z(\beta) = 2^{N N_F} \int \Big({\textstyle \prod_i d x_i} \Big) \, e^{- N^2 S[\{x_i\}]} \,.
\ee
It is natural to keep the factor of $2^{N N_F}$ explicit to connect to the original sum over $N N_F$ spins in (\ref{eq:original}).
Note that we are dealing with an integral over real symmetric matrices rather than hermitian matrices, this will affect the coefficient of the repulsive inter-eigenvalue potential.
Working to leading order at large $N$ we can use the expression
\be
\text{vol}_{SO(N)} \sim e^{3 N^2/8} \left(\frac{4 \pi}{N}\right)^{N^2/4} \,,
\ee
and obtain the eigenvalue action
\be\label{eq:eaction}
S[\{x_i\}] =  \frac{1}{N} \sum_i \left[- \frac{1}{2 N} \sum_{j \neq i} \log \frac{|x_i - x_j|}{\beta^{1/2}} +  \frac{1}{\beta} x_i^2 + \frac{\alpha}{2} \log \left(\gamma -  x_i \right) \right] - c  \,.
\ee
The constant
\be
c = \frac{3 + 4 \log 2}{8} + \frac{\a (\g - 1)}{2} \,.
\ee
The equations of motion following from (\ref{eq:eaction}) are
\be\label{eq:eom}
\frac{1}{N} \sum_{j\neq i} \frac{1}{x_i - x_j} = \frac{2}{\beta} \, x_i - \frac{1}{2} \frac{\alpha}{\gamma -  x_i} \,.
\ee
Introducing the eigenvalue distribution (normalized to unity)
\be
\rho(y) = \frac{1}{N} \sum_i \delta(y - y_i) \,,
\ee
then in the large $N$ limit the equations of motion (\ref{eq:eom}) become
\be\label{eq:continuum}
\beta \, P\int \frac{dy \rho(y)}{x-y} = 2 \, x - \frac{1}{2} \frac{\a \, \beta}{\gamma - x} \,.
\ee

The integral equation (\ref{eq:continuum}) can be solved for the eigenvalue distribution using standard techniques \cite{Brezin:1977sv}. The following solution is found
\be\label{eq:AB}
\rho(y) =  \frac{2}{\pi \beta} \left(1 - \frac{a+b}{2(\gamma-y)} \right) \sqrt{(y-a)(b-y)} \,, \qquad a \leq y \leq b < \gamma \,,
\ee
with vanishing support elsewhere.
\begin{figure}[t]
\begin{center}
\includegraphics[height = 65mm]{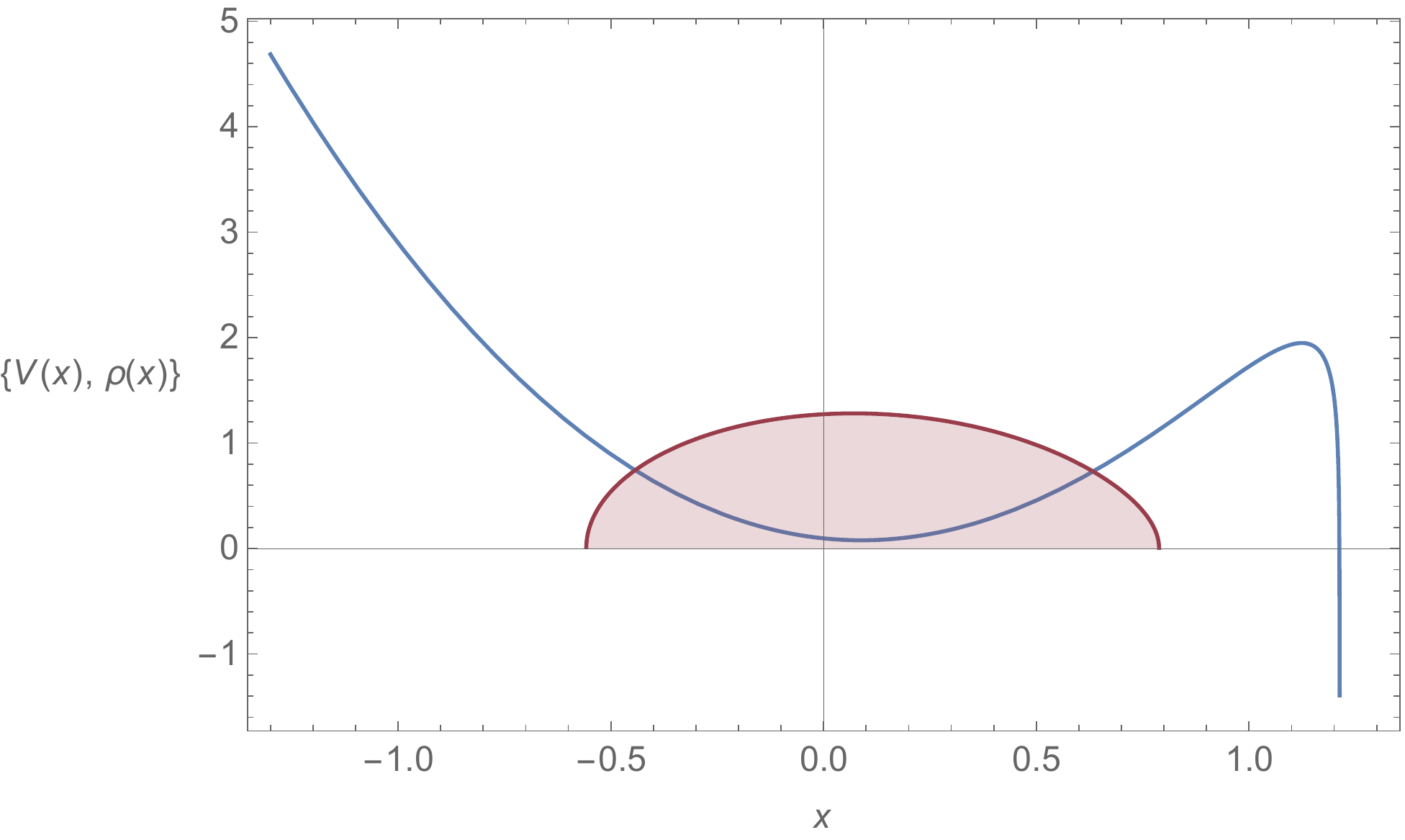}\caption{The eigenvalue distribution at $T = 1/\beta = 2.5$ and $\alpha = 1$. Also shown is the external potential (\ref{eq:V}) experienced by the eigenvalues. \label{fig:exist}}
\end{center}
\end{figure}
The endpoints $a$ and $b$ must satisfy two conditions
\bea
\sqrt{(\g-a)(\g-b)} (a+b) & = & \frac{\alpha \, \beta}{2} \,, \label{eq:AB1} \\
\frac{3 b^2 + 2 a b + 3 a^2 - 3 \beta + 2 \alpha \beta - 4(a + b)\gamma}{\beta} & = & 1 \,.\label{eq:AB2}
\eea
The first of the above two equations ensures that the final term on the right hand side of (\ref{eq:continuum}) is matched, while the second is the normalization condition. Finally, the constraint (\ref{eq:constraint}) becomes
\be
\int \frac{\rho(x) dx}{\g - x} = \frac{(b-a)^2 (a+b)+2 ([a+b]^2 + \a \beta) \g - 4(a+b) \g^2}{\a \beta^2} = 1 \label{eq:AB3} \,.
\ee
For a given $\a,\beta$, the eigenvalue distribution is fully specified once we have solved (\ref{eq:AB1}) -- (\ref{eq:AB3}) for $a,b$ and $\gamma$. The eigenvalue distribution must be nonnegative everywhere.
An example eigenvalue distribution is plotted in figure \ref{fig:exist}, together with the external potential appearing in the eigenvalue action (\ref{eq:eaction}):
\be\label{eq:V}
V_\text{ext.}(x) = \frac{1}{\beta} x^2 + \frac{\alpha}{2} \log \left(\gamma -  x \right) \,.
\ee

In figure \ref{fig:exist} we see the first indication that the matrix integral saddle point we have found may be metastable: the potential becomes unbounded below at $x = \gamma$. Furthermore, as the temperature is lowered (i.e. $\beta$ large), then the repulsive logarithmic term becomes more important in the external potential (\ref{eq:V}). Eventually, the repulsion between eigenvalues will push them over the maximum seen in figure \ref{fig:exist} and the solution will no longer exist. This is a well known phenomenon \cite{Brezin:1977sv}. Indeed, it is found that physical solutions to the algebraic equations (\ref{eq:AB1}) -- (\ref{eq:AB3}) only exist in the range of temperatures
\be\label{eq:range}
0 \leq \beta \leq \beta_c \,.
\ee
A plot of the critical temperature as a function of $\alpha$ is shown in the following figure \ref{fig:Tcrit}.
\begin{figure}[ht]
\begin{center}
\includegraphics[height = 60mm]{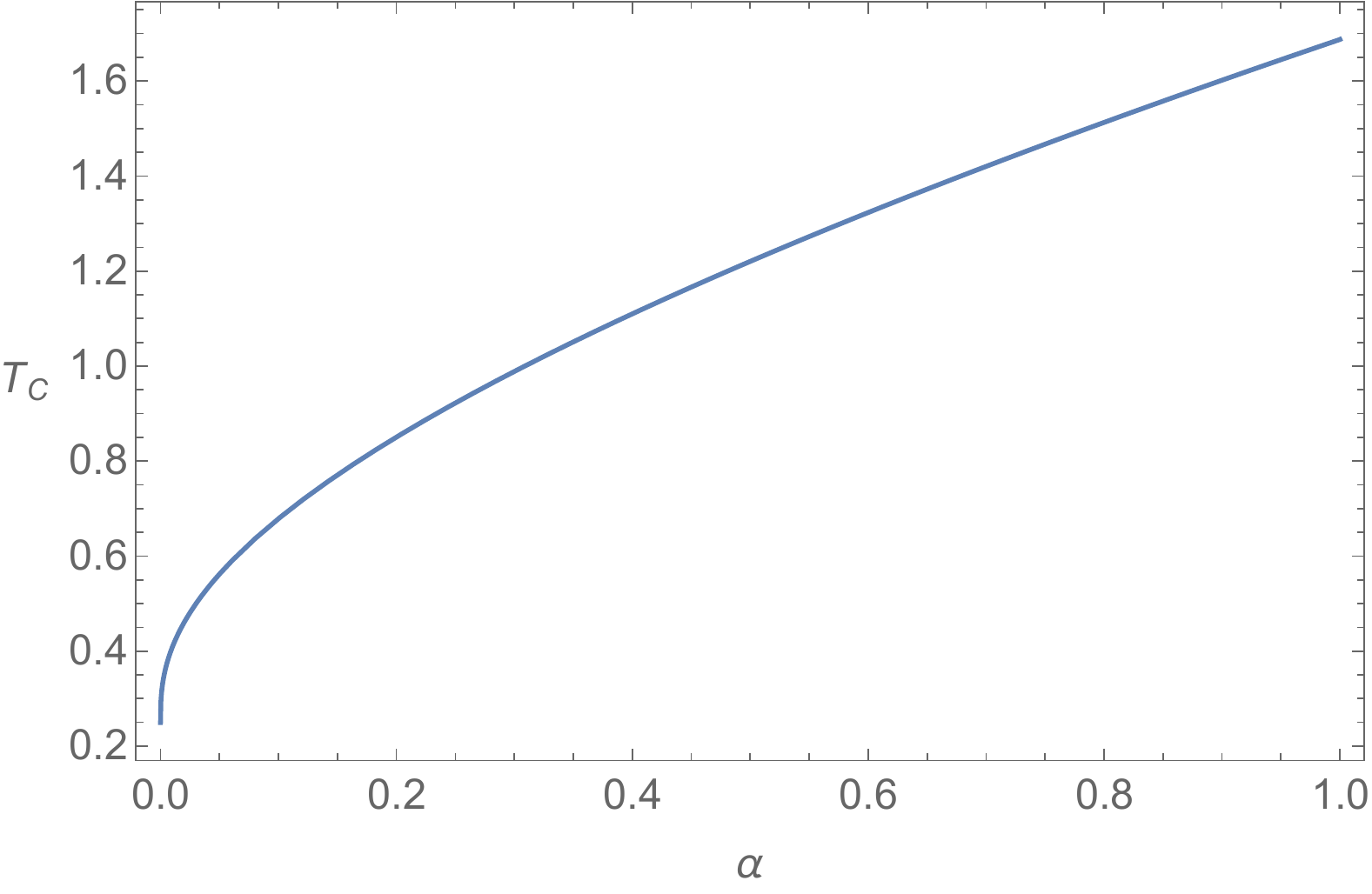}\caption{Temperature below which there is no eigenvalue saddle, as a function of $\alpha \in [0,1]$.\label{fig:Tcrit}}
\end{center}
\end{figure}
The limiting values are
\be\label{eq:criticalT}
\lim_{\alpha \to 0^+} T_c = \frac{1}{4} \,, \qquad \lim_{\alpha \to 1} T_c = \frac{27}{16} \,.
\ee
As we might have anticipated, smaller $\alpha$ allows the saddle point to exist to lower temperatures, as it reduces the strength of the logarithmic term.\footnote{When $\alpha$ exactly vanishes, there is no critical temperature -- the distribution in this case is just the Wigner semicircle distribution at all temperatures. Thus the limit $\alpha \to 0$ is discontinuous -- at any nonzero $\alpha$ the saddle disappears below an order one temperature.} At the critical temperature we find
\be\label{eq:silly}
1 - \frac{a+b}{2(\g - b)} = 0\,,
\ee
(thus, for instance, with $\alpha=1$ and at the critical temperature: $\gamma = \frac{4}{3}, a = - \frac{2}{3}, b = \frac{10}{9}$)
and hence the distribution (\ref{eq:AB}) is on the verge of becoming negative near the upper bound $x =b$. The critical distribution at $\beta = \beta_c$ therefore vanishes like $(b-x)^{3/2}$ rather than $(b-x)^{1/2}$ as $x \to b$. This is a familiar effect for these types of transition \cite{Brezin:1977sv}.

The derivation of the matrix integral from the full partition function, in particular the $z(Q)$ part (\ref{eq:zqexpand}), gives a `physical' interpretation of the instability. The $w$ vector modes have been integrated out to obtain the matrix model (\ref{eq:Zfinal}) for $Q$. However, once the eigenvalue distribution for $Q$ is pushed towards $x = \gamma$, then some of these modes that were integrated out become massless. One must then contend with the whole partition function. Indeed, we will see in section \ref{sec:meta} below that at low temperatures the partition function is dominated by $Q$ matrices with a large eigenvalue and that these configurations compete with the matrix integral saddle discussed so far.

At high temperatures, the distribution becomes narrow as the quadratic term in the action (\ref{eq:eaction}) dominates. The physical solution to the algebraic equations behaves as
\be
a = - \beta^{1/2} + \frac{\alpha \, \beta}{4} + \cdots \,, \quad
b = \beta^{1/2} + \frac{\alpha \, \beta}{4}  + \cdots \,, \quad
\gamma = 1 + \frac{(1+\alpha) \beta}{4} + \cdots \,.
 \label{eq:highT}
\ee
This expansion allows us to evaluate the on shell action in a high temperature expansion. In the continuum large $N$ limit the action (\ref{eq:eaction}) becomes
\be\label{eq:actcont}
S = - \frac{1}{2} \int dxdy \rho(x) \rho(y) \log \frac{|x-y|}{\beta^{1/2}} + \int dx \rho(x) \left(\frac{x^2}{\beta} + \frac{\alpha}{2} \log(\g-x) \right)  - c \,. 
\ee
To evaluate on shell, we may use the fact that for distributions satisfying the equations of motion (\ref{eq:continuum}) we have
\be
\int dy \rho(y) \log \frac{|x-y|}{\beta^{1/2}} = \frac{x^2}{\beta} + \frac{\alpha}{2} \log(\gamma-x) + \kappa \,.
\ee
The constant $\kappa$ is easily fixed by evaluating the above expression at e.g. the upper endpoint $x = b$. All the integrals necessary to evaluate the action can be done explicitly. The difficulty for obtaining an explicit answer is the need to solve the algebraic equations (\ref{eq:AB1}) -- (\ref{eq:AB3}) for $a,b,\g$. In a high temperature expansion  (\ref{eq:highT}), the on shell action is found to be
\be\label{eq:highTact}
S  = - \left( \frac{(1 + \a) \a \beta}{16} + \frac{\a^2 \beta^2}{128} + \frac{\a^2 (1+\a) \beta^3}{768} + \frac{\a^2(1 + 4 \a + \a^2) \beta^4}{4096} + \cdots \right) \,.
\ee
These coefficients are essentially counting the number of contractions that contribute to the free energy at order $N^2$ in an expansion of the spin partition function (\ref{eq:original}). In fact, the first term is simply the Hamiltonian averaged over all spin configurations times $\beta/N^2$. To see this first note that the terms in the Hamiltonian that involve four different spins average to zero. However, the terms that involve only two different spins, i.e. $\sigma_{Ab}^2\sigma_{Ad}^2$ and $\sigma_{Ab}^2 \sigma_{Cb}^2$, are identically equal to one. There are $N^2 N_F+N N_F^2$ of these terms, which leads to an average energy $E=- N^2 (1+\a)\a/16$. Using that $E=N^2 (\partial S/\partial \beta)$ we verify the first term in (\ref{eq:highTact}). 

In the case $\alpha =1$, where $N = N_F$, we have been able to obtain the action in closed form (up to the need to solve for $\gamma$). If we define
\be
\bar \beta = \frac{\beta}{\gamma^2} \,, \qquad \bar c = \frac{1-\g + \log \g}{2} \,,
\ee
then the on shell action is found to be
\bea \label{eq:actionalpha1}
S & = & - \sum_{n=1}^\infty \frac{(2n-1)!}{(n+2)! \, n!} \left(\frac{3 \bar \beta}{4}\right)^n + \bar c \label{eq:ShighT}\\
& = & - \frac{\bar \beta}{8} {}_3 F_2 \left(1,1,\frac{3}{2}; 2,4; 3 \bar \beta \right) + \bar c \\
&= & \frac{2 - 15 \bar \beta}{54 \, \bar \beta^2} \sqrt{1 - 3 \bar \beta} + \frac{1}{2} \log \frac{1 + \sqrt{1 - 3 \bar \beta}}{2} + \frac{-8+72 \bar \beta - 81 \bar \beta^2}{216 \bar \beta^2} + \bar c \,. \label{eq:sexact}
\eea
We originally found this result by inspection of the high temperature expansion for the action (\ref{eq:highTact}) to high order. We have also used the fact that $\gamma$ can be scaled out of the on shell action (\ref{eq:actcont}) if $\beta$ is also rescaled to $\bar \beta$. It turns out that (\ref{eq:ShighT}) is very closely related to a familiar combinatorial expansion \cite{math1, math2} that can be exactly resummed. The discussion in \cite{math1, math2} concerns a quartic rather than our logarithmic interaction between the matrices. 
The connection between the two is due to the fact that a quartic reformulation of the constrained logarithmic problem is possible when $\alpha = 1$. This quartic formulation was used in \cite{parisi}; we adapt their computation to our ferromagnetic case in appendix \ref{sec:quartic} and recover several of our results from this alternative perspective.

The full expression (\ref{eq:sexact}) allows us to see easily the non-analyticity appearing as $\bar \beta \to \bar \beta_c = 1/3$. We noted below (\ref{eq:silly}) above that at the critical temperature $\gamma = \frac{4}{3}$ for $\alpha =1$, and hence we recover the critical temperature $\beta_c = \gamma^2 \bar \beta_c = \frac{16}{27}$ that we quoted above in (\ref{eq:criticalT}). The non-analyticity appears as
\be
S = \bar c + \frac{7 - 48 \log 2}{24} + \frac{1}{4} (\bar \beta_c - \bar \beta) - \frac{9}{8} (\bar \beta_c - \bar \beta)^2 - \frac{12 \sqrt{3}}{5} (\bar \beta_c - \bar \beta)^{5/2} + \cdots \,.
\ee
This corresponds to a very weak non-analyticity in the free energy
\be
F = - T \log Z = N^2 T S \,.
\ee
We expect there to be a first order transition at $T_c$, as we will confirm shortly, and so this weak non-analyticity will only be seen coming from the high temperature side of the transition. Furthermore, we will see that the eigenvalue saddle is metastable (but long lived as $N \to \infty$) by the time it disappears at $T_c$.

While we have presented a closed-form answer for the free energy only for $\alpha = 1$, in the following subsection we will derive an exact expression for the energy
\be
E = F - T \frac{dF}{dT} = N^2 \frac{d S}{d \beta} \,,\label{eq:ds}
\ee
for all $\alpha$.

\subsection{Exact expression for the energy}

From the definition of the energy in (\ref{eq:ds}), together with the matrix integral partition function (\ref{eq:Zx}), we have
\bea
\frac{E}{N^2} = - \frac{1}{N^2} \frac{1}{Z} \frac{d Z}{d\beta} & = & \int dx \rho(x) \left( - \frac{x^2}{\beta^2} + \frac{1}{4\beta} \right)  \\
& = & \frac{1 - \alpha  \gamma}{4 \beta } -\frac{(b-a)^2}{64 \beta^3} \Big( (3 b+a) (b+3 a)-4 (\alpha -2) \beta \Big) \,. \label{eq:enexact}
\eea
Here the eigenvalue distribution is the solution (\ref{eq:AB}). The integral in the first line may be performed exactly.
To obtain the final line we have simplified the expression slightly using the equations (\ref{eq:AB1}) -- (\ref{eq:AB3}) for $a,b,\g$. We need to solve those algebraic equations numerically to obtain a numerical value for the energy from (\ref{eq:enexact}). For the case of $\alpha = 1$, the equations for $a$ and $b$ can be solved analytically. The energy then agrees with the result following from (\ref{eq:sexact}).

\subsection{Verification by Monte Carlo numerics}

We have verified the analytic results above by performing Monte Carlo numerics on the original spin system (\ref{eq:isingH}). We used a simple Metropolis algorithm. For concreteness we have focussed on the $\alpha = 1$ case. The results presented below used a $200 \times 200$ system size. We ran several tens of Monte Carlo sweeps at each temperature. For each temperature we made a histogram of all of the values of the energy obtained over the many Monte Carlo time steps (neglecting the steps in the first sweep). After enough sweeps were performed, this gave us an approximately Gaussian distribution of energies, from which we extracted the average value as well as the standard deviation. This deviation is intrinsic, due to the finite size of the system, and eventually does not decrease upon running for a longer time. The (within the error bars) systematic error in the data is due to terms subleading in $N$ that are neglected in the matrix integral result.

Figure \ref{fig:MC} below shows the results of the Monte Carlo simulations together with the matrix saddle result (\ref{eq:enexact}) for the energy. The plot is concerned with $T > T_c$. The agreement seems to be compelling and
\begin{figure}[h]
\begin{center}
\includegraphics[height = 80mm]{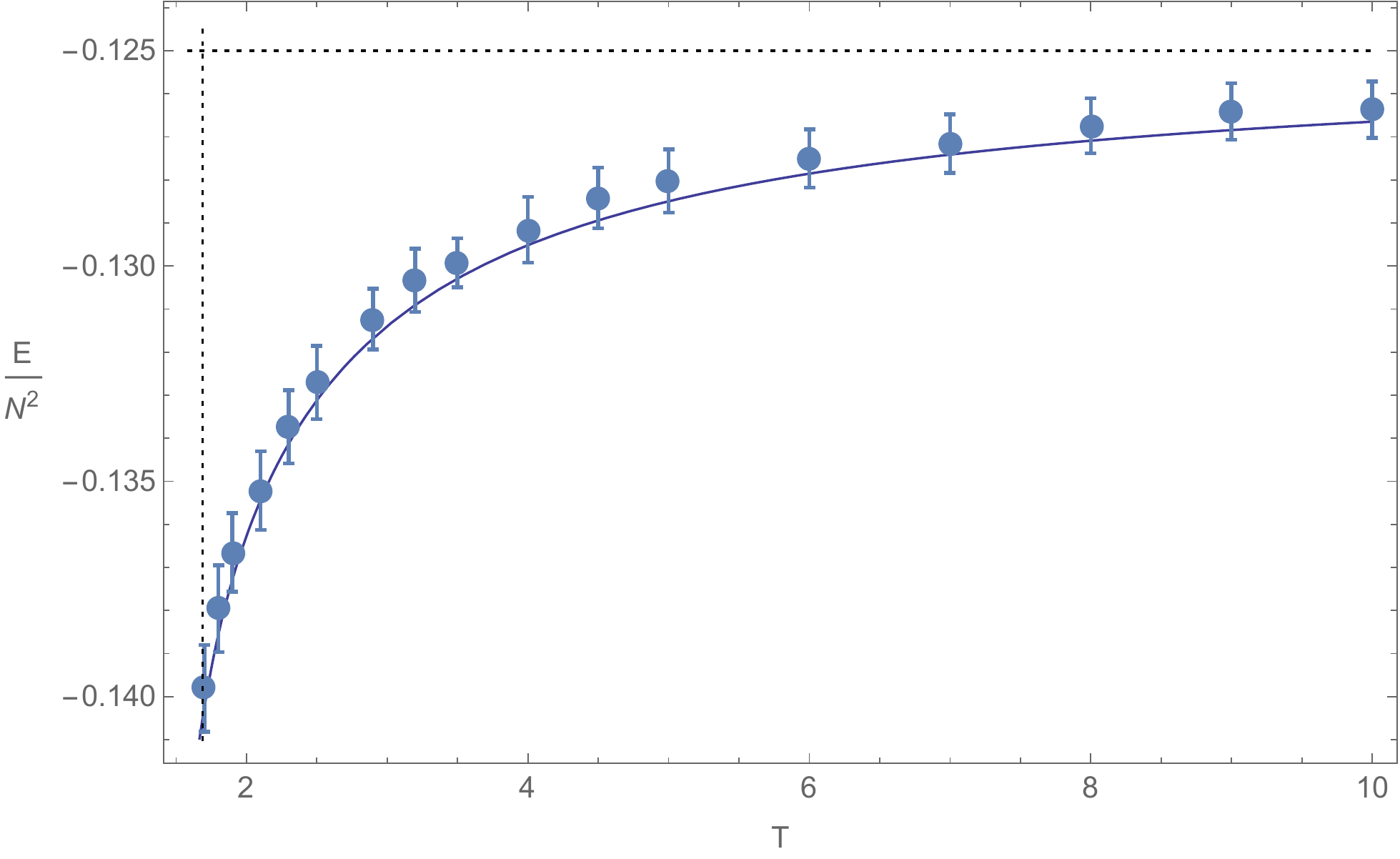}\caption{Energy versus temperature for $\alpha = 1$ (with $N = N_F = 200$) at temperatures above $T_c$. Dots are Monte Carlo data points with one sigma error bars. The solid line is the matrix integral result following from (\ref{eq:enexact}). The dotted lines show the critical temperature $T_c = 27/16$ and the asymptotic energy $E = - N^2/8$.\label{fig:MC}}
\end{center}
\end{figure}
validates our conclusion that for $T > T_c$ the nonlocal spin system (\ref{eq:isingH}) is indeed described by the matrix integral (\ref{eq:Zfinal}). We will see shortly that, in fact, for temperatures below a much higher temperature, $T_* \sim N$, the matrix integral saddle is metastable. Nonetheless, because the Monte Carlo works by cooling from a high temperature configuration, it gets trapped in the metastable configuration for a very long time. As $N \to \infty$, the metastable configuration becomes infinitely long lived and so is physically meaningful. We can see this from the eigenvalue potential (\ref{eq:V}), plotted in figure \ref{fig:exist}. The eigenvalue distribution is locally stabilized by a free energy barrier of height $N^2$.

For $T < T_c$ the Monte Carlo converges to one of the ground states of the system. The transition at $T = T_c$ therefore appears to be a first order transition to a non-matrix integral saddle. We turn now to a discussion of this transition and of ground states.

\subsection{Ground states and metastability}
\label{sec:meta}

The ferromagnetic ($\lambda = - \frac{1}{16 N} < 0$) nonlocal Ising model (\ref{eq:isingH}) has a large number of ground states that are easily characterized. The first is when all spins are up: $\sigma^{Ab} = 1$ for all $A,b$. The remaining ground states are all obtained from this one by acting with the symmetries described above equation (\ref{eq:isingH}). The symmetry operations are flipping all the spins in a single row or column, and permutations of rows or columns. The ground state energy and degeneracy are thus seen to be
\be\label{eq:gs}
E_0 = - \frac{N N_F^2}{16} \,, \qquad g_0 = 2^{N+N_F - 1} \,.
\ee
The reason the degeneracy is not $2^{N+N_F}$ is that if we flip the sign of all the rows and columns then we get back the same state we started from. While the ground state degeneracy is large, it is not extensive. The possibility of characterizing the ground states explicitly is one difference between the ferromagnetic and the antiferromagnetic cases. The antiferromagnetic case exhibits frustration and glassiness at low temperatures \cite{parisi}. On the other hand, our ferromagnetic case has ordered low energy states with energies $E \sim -N^3$. These ordered states will necessarily dominate at low temperatures. 

In the Monte Carlo simulation we indeed observe that below a certain temperature one of the ground states is reached after some number of sweeps. The sudden drop in energy clearly indicates a first order phase transition. To determine the transition temperature accurately a different method is needed. One should look at the canonical distribution
\be
P(E,T) = g(E) e^{-\beta E},
\ee
where $g(E)$ is the density of states. For first order phase transitions, the canonical distribution has a characteristic doubly peaked structure as a function of energy near the transition temperature. For finite $N$ the transition temperature is determined by the temperature at which the two peaks are of equal height. Using a Monte Carlo simulation to obtain the density of states, we can obtain a fairly accurate value for the transition temperature \cite{MCdos}. This simulation is temperature independent and therefore does not suffer from hysteresis or thermal fluctuations to the metastable state. The Monte Carlo simulations for the density of states converge much more slowly than the simulations at fixed temperature. This limits the accessible system sizes significantly. Nevertheless, already for small systems, we clearly observe the doubly peaked structure of the canonical distribution. See figure \ref{fig:MCdos} below.
\begin{figure}[h]
\begin{center}
\includegraphics[height = 70mm]{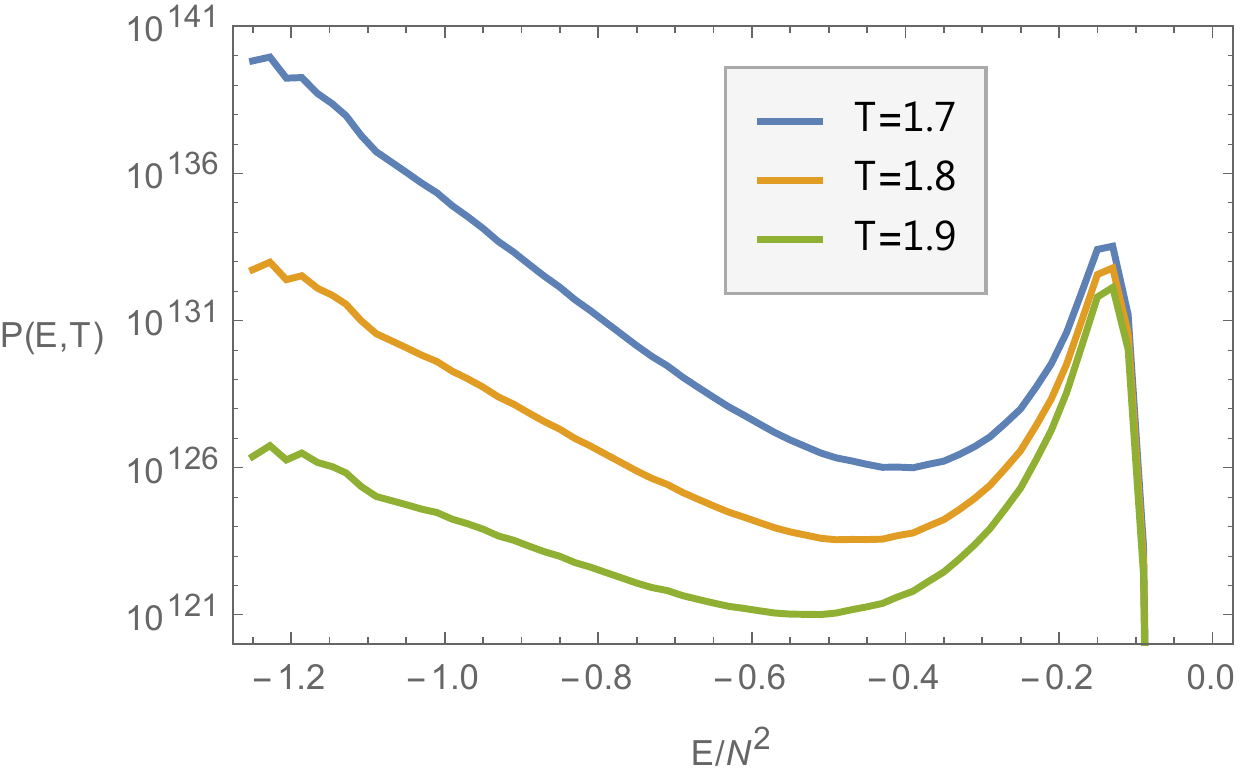}
\caption{Canonical distribution $P(E,T)$ as a function of energy at three temperatures around $T_* \approx 1.8$ for $N=N_F=20$. We smoothened the data by integrating the density of states on energy intervals $\Delta E/N^2=0.02$. Above $T_*$ the peak around $E=-N^2/8$ is highest, whereas below $T_*$ the peak at $E=-N^3/16$ is highest.\label{fig:MCdos}}
\end{center}
\end{figure}

Inspection of the density of states, e.g. in figure \ref{fig:MCdos}, reveals that there are two classes of spin configurations. For the $N$ counting, recall $N_F \sim N$. Firstly, there are order $2^N$ states with energies close to the ground state energy, so that $E \sim - N^3$. These are the ground states themselves together with some flipped spins. Then there are order $2^{N^2}$ states whose energies are of order $E \sim - N^2$. These are the configurations that contribute to the matrix integral saddle point. These two contributions will be roughly the two peaks in the canonical distribution. Thus, very very schematically, we may write the partition function as
\be\label{eq:Zrough}
Z \sim 2^N e^{N^3/T} + 2^{N^2} e^{N^2/T} \,.
\ee
This simplified model of the partition function captures a key non-commutativity of limits. If we keep $N$ fixed and take the high temperature limit, then for temperatures above $T_* \sim N$ the matrix integral saddle (the second term) dominates because of its higher entropy (which is of order $N^2$). However if we keep $T$ fixed and take the large $N$ limit, then the low lying states always dominate the partition function because their energy is much lower. Recall that we derived the matrix description (\ref{eq:Zfinal}) of the partition function in a formal high temperature expansion to all orders. This non-commutativity of limits explains why the matrix saddle can be metastable or disappear entirely at low temperatures. The competition between the matrix configurations and the ground states is one of entropy versus energy, as is characteristic for first order phase transitions.

\begin{figure}[ht]
\begin{center}
\includegraphics[height = 80mm]{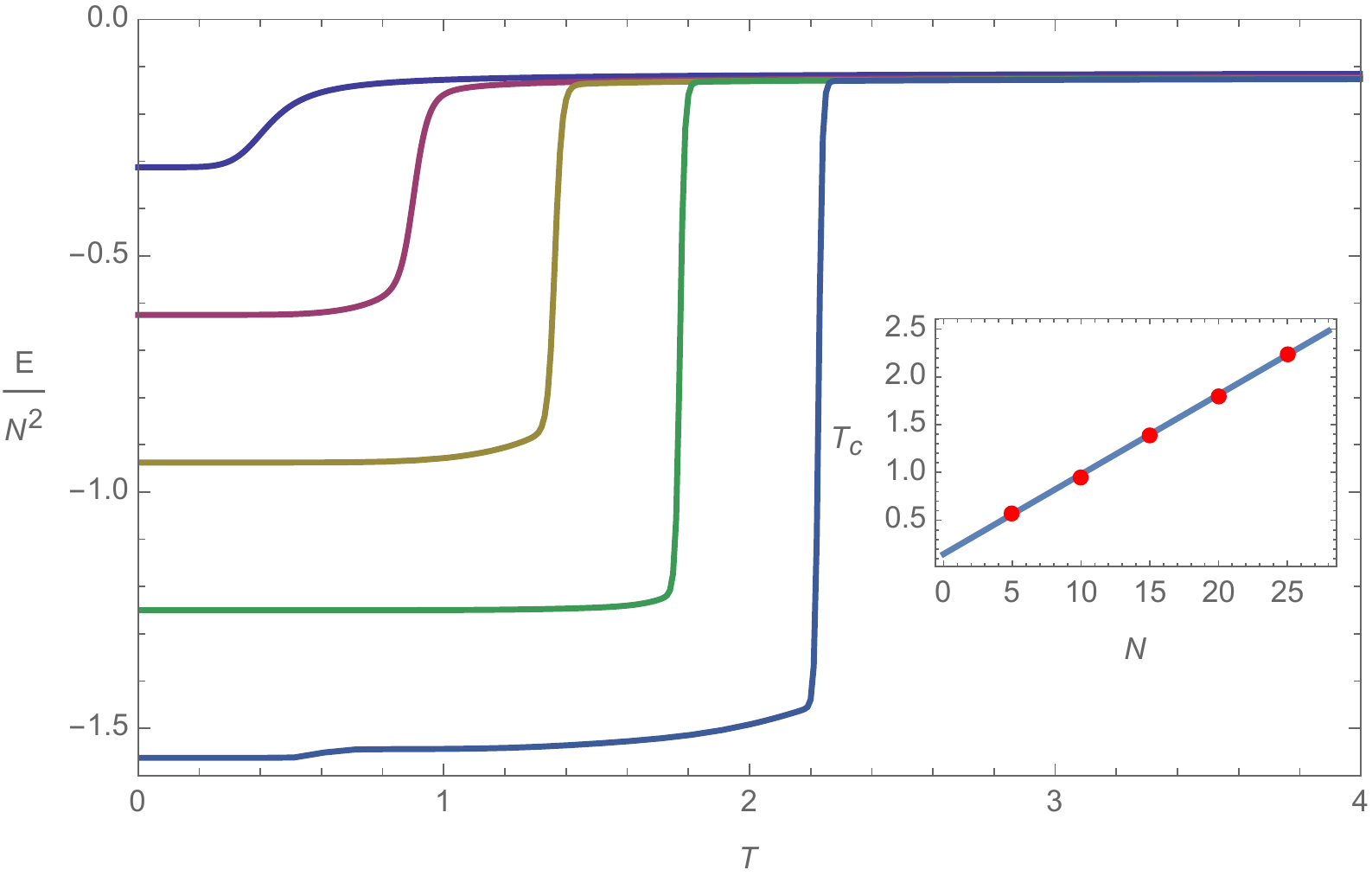}
\caption{The energy expectation value, $E=\sum_E E P(E,T)/Z$, as a function of temperature for $N=5,10,15,20,25$ (top to bottom), showing the sharp drop to the ground state energy $E=-N^3/16$ below $T_*$. The inset shows $T_*$ as a function of $N$ determined from the temperature where the peaks of $P(E,T)$ are of equal height. The line is a linear fit to the data: $T_* \approx 0.08 N + 0.15$. \label{fig:MCdos2}}
\end{center}
\end{figure}

Varying $N$ in the numerics, we find a linear dependence of the transition temperature $T_*$ on $N$. In figure  \ref{fig:MCdos2} we see -- using low values of $N=N_F=5,10,15,20,25$ --  that the peaks in plots such as figure \ref{fig:MCdos} exchange dominance at
\be\label{eq:TN}
T_* \approx 0.08  N \,. \qquad  \qquad (\text{for } N=N_F)
\ee
This value can be understood using the logic of (\ref{eq:Zrough}) at high temperatures: The ground state energy (\ref{eq:gs}) and the degeneracy of the matrix states (\ref{eq:Zx}) become equal at $T_* = N/(16 \log 2) \approx 0.09 N$, in good agreement with (\ref{eq:TN}), given that the data is not fully in the large $N$ limit. At large $N$ this temperature is much greater than the $T_c$ at which the matrix integral saddle ceases to exist.

Finally, we note that the structure of the partition function in (\ref{eq:Zrough}) indicates an easy way that the model can be modified to render the matrix model saddle point stable rather than metastable over the range of temperatures that it exists. It is necessary to modify the Hamiltonian (\ref{eq:isingH}) in such a way that the states with energy of order $-N^3$ are lifted while the states with energy $-N^2$ are untouched. One simple way to do this is to shift the Hamiltonian $H \to H + \frac{c^2}{N^3} H^2$, for some suitably large constant $c$. This shift sends the states with energy $-N^3$ to energy $+N^3$ while having negligible effect on the states with energy $N^2$. The following section \ref{sec:largeQ} will lead to a general picture of what is necessary to stabilize the matrix integral saddle: the contribution of matrices $Q$ with large eigenvalues must be suppressed in the partition function (\ref{eq:third}).

To put everything together, it is instructive to understand where the ground state contribution is to be found when the partition function is expressed as an integral over the $Q$ matrix (\ref{eq:third}). We have already seen that the matrix saddle involves $Q$ matrices with eigenvalues or order one or less. In the following subsection we see that the first term in (\ref{eq:Zrough}), the `ground state' contribution, comes from $Q$ matrices with large eigenvalues.

\subsection{Low temperature expansion and large eigenvalues}
\label{sec:largeQ}

Consider a low temperature expansion of the partition function (\ref{eq:third}), written as an integral over the $Q$ matrix. As $\beta \to \infty$ the integral is dominated by large $Q$. It follows that in this limit the reduced partition function $z(Q)$ is dominated by its ground states. This is because the overall magnitude of $Q$ can be considered as an inverse temperature in the reduced partition function (\ref{eq:reduced}). Therefore we have
\be
z(Q) \to g_0(Q) e^{-E_0(Q)} \,,
\ee
where $g_0(Q)$ and $E_0(Q)$ are the ground state degeneracy and energy of the Hamiltonian 
\be\label{eq:reducedh}
h = -\frac{1}{2} Q_{AB} \sigma^A \sigma^B \,.
\ee
The full partition function becomes
\be
Z(\beta) \stackrel{\beta\to\infty}{=} \frac{1}{2^{N/2}} \left(\frac{2 N}{\pi\beta}\right)^{N(N+1)/4}  \int dQ e^{- \frac{N}{\beta} Q_{AB} Q_{AB} - N_F E_0(Q) + N_F \log g_0 (Q)} \,.
\ee

In the above expression we see that the matrices $Q$ that lead to the lowest ground state energies $E_0(Q)$ (at fixed $Q_{AB} Q_{AB}$) will give the largest contribution to the integral. Consider the set of matrices where we fix the magnitudes of the components of the matrix to be $|Q_{AB}|$, but allow arbitrary signs $\pm$. Among this set, the ground state energy of the spin system (\ref{eq:reducedh}) is minimized by the matrix with all positive signs {\it and} by all matrices related to this matrix by changing the sign of any number of rows and corresponding columns (the $Q$ matrix is symmetric). The ground state energy for these $2^{N-1}$ matrices is\footnote{Note that the number of symmetric matrices related to (and including) a completely positive matrix by simultaneously flipping the sign of rows and corresponding columns is $2^{N-1}$. The na\"ive answer that there are $2^N$ such matrices counts each matrix twice, since the matrix obtained by flipping a set of rows and corresponding columns can also be obtained by flipping all the \emph{other} rows and corresponding columns.}
\be
E_0(Q) = - \frac{1}{2} \sum_{A, B} |Q_{AB}|\,,
\ee
and the degeneracy $g_0(Q) = 2$ corresponding to flipping all the spins. The contribution of this set of matrices to the partition function is then
\be\label{eq:positive}
Z_+(\beta) \stackrel{\beta\to\infty}{=}  \frac{1}{2^{N/2}} \left(\frac{2 N}{\pi\beta}\right)^{N(N+1)/4} 2^{N_F+ N-1} \int dQ_+ e^{- N \sum_{A,B} \left( \frac{1}{\beta} Q_{AB} Q_{AB} - \frac{\alpha}{2} Q_{AB} \right)} \,.
\ee
Here the $+$ index indicates that we have restricted the integral to being over matrices with only positive entries. We have argued that these should dominate the low temperature partition function. The factor of $2^{N-1}$ describes the contributions from matrices related to these by flipping the signs of rows and columns as described above. It is easy to do the integrals in (\ref{eq:positive}) by completing the square and using the large $\beta$ limit to write $\int_{-\alpha \beta/4}^{\infty} f(x) dx$ as $\int_{- \infty}^{\infty} f(x) dx$, giving
\be
Z_+(\beta) \stackrel{\beta\to\infty}{=} 2^{N_F + N-1} e^{\frac{\beta}{16} N N_F^2} \,.
\ee
which is precisely the ground state contribution expected from the energies and degeneracies given in (\ref{eq:gs}) above. We have made no large $N$ approximation in this computation.

The above computation (see equation (\ref{eq:positive})) shows that the ground state contribution to the partition function comes from matrices with entries
\be\label{eq:largeQavg}
Q_{AB} \sim \frac{\a \beta}{4} \,.
\ee
In particular, the matrix will have one very large eigenvalue, of order $N_F \beta$.
As anticipated, this is distinct from matrices contributing to the matrix integral saddle
which had all eigenvalues of order one or smaller.

In fact, at large $N$, the shift $\int_{-\alpha \beta/4}^{\infty} f(x) dx \to \int_{- \infty}^{\infty} f(x) dx$ can be performed at any temperature because the Gaussian $f(x) = e^{- N x^2/\beta}$ has a very narrow width. Therefore, the computations above
extract the large $Q$ contribution to the partition function at any temperature, so long as $N$ is large. Even away from high temperatures, the competition between the entropic and energetic contributions to the partition function (\ref{eq:Zrough}) appears as the competition between $Q$ matrices with small and large eigenvalues.

\subsection{Order and disorder}\label{sec:defects}

In this section we argue that the low temperature saddle describes an ordered phase, while the matrix integral saddle corresponds to a disordered phase. Remember that all the ground states are related to the state with all spins up by the 
symmetries that flip the spins along rows and columns. The ground state energy is $E_0=- N N_F^2/16$ and follows from the fact that each term in the Hamiltonian is $-1/(16N)$ for these states. Na\"ively, the lowest excited states have one spin flipped 
compared to a ground state. \begin{figure}[ht]
\begin{center}
\includegraphics[height = 65mm]{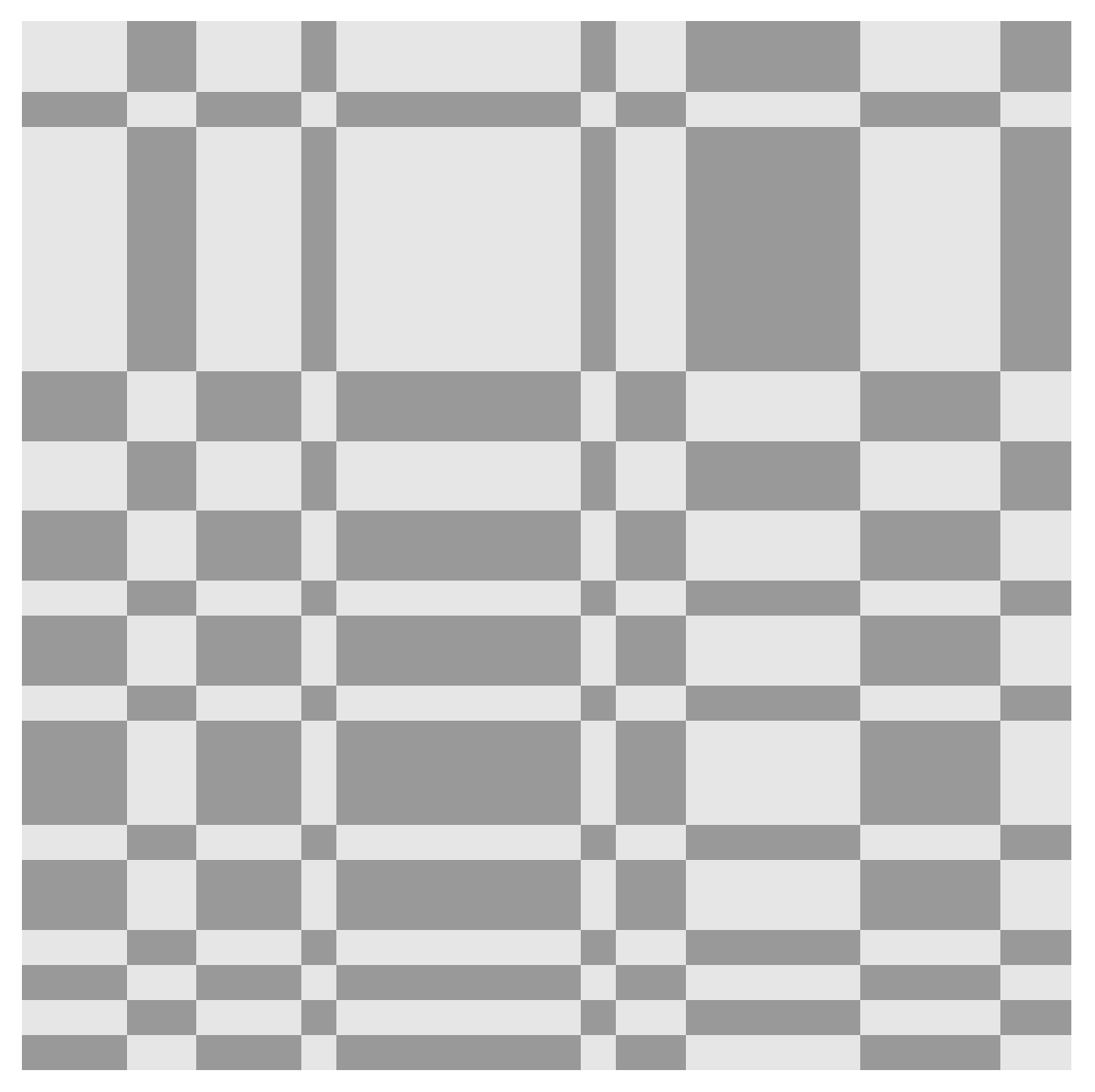}
\includegraphics[height = 65mm]{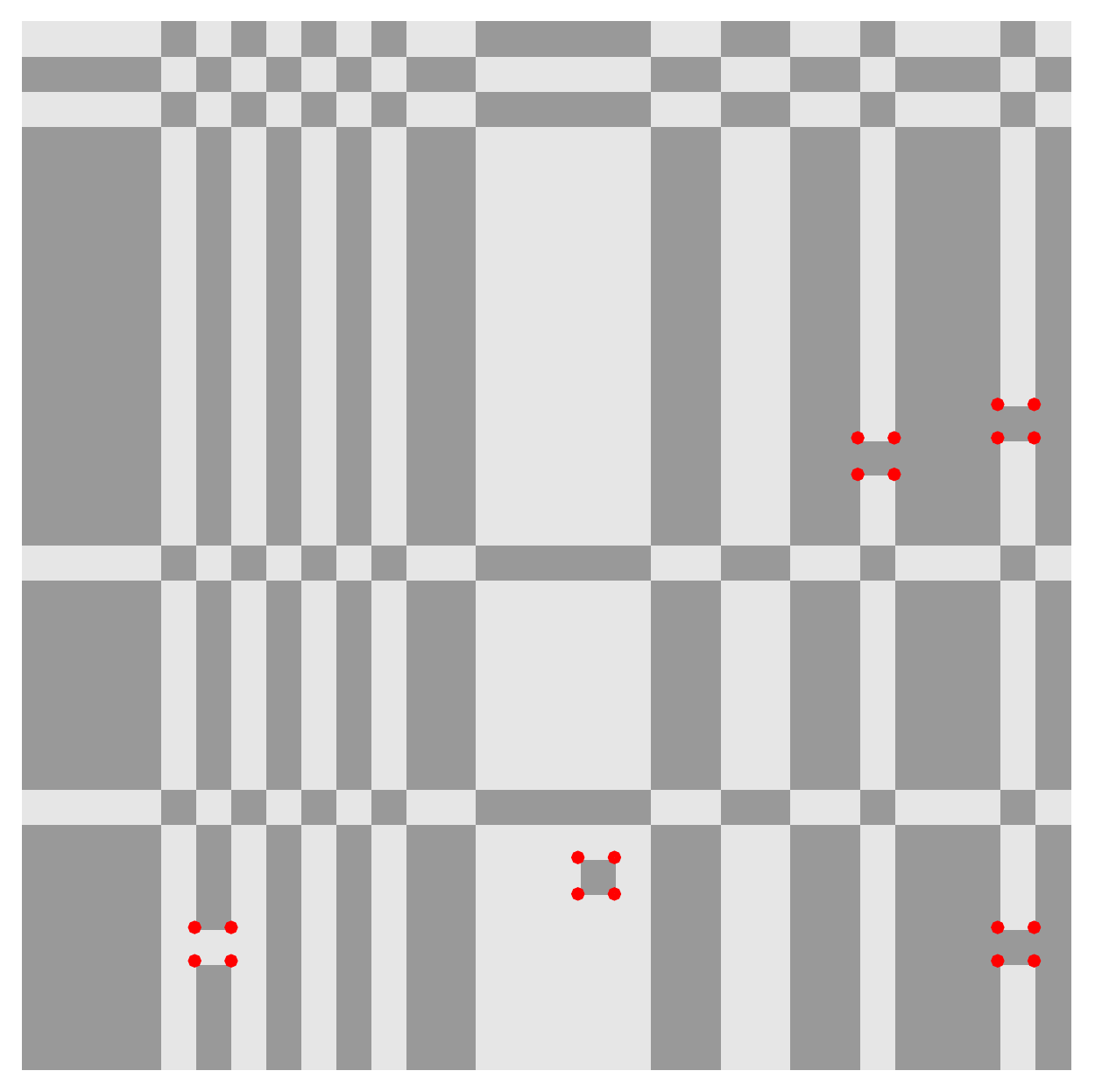}
\includegraphics[height = 65mm]{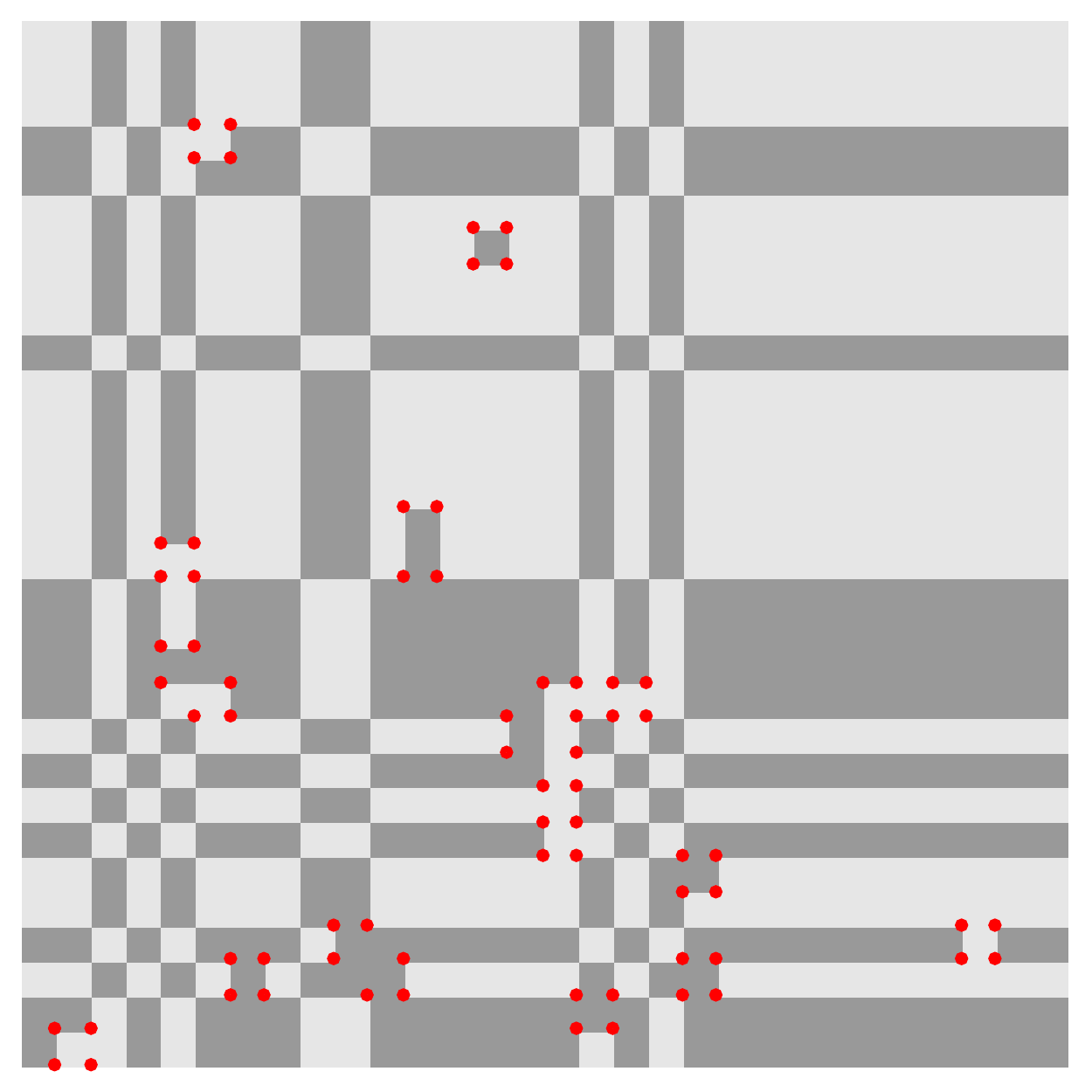}
\includegraphics[height = 65mm]{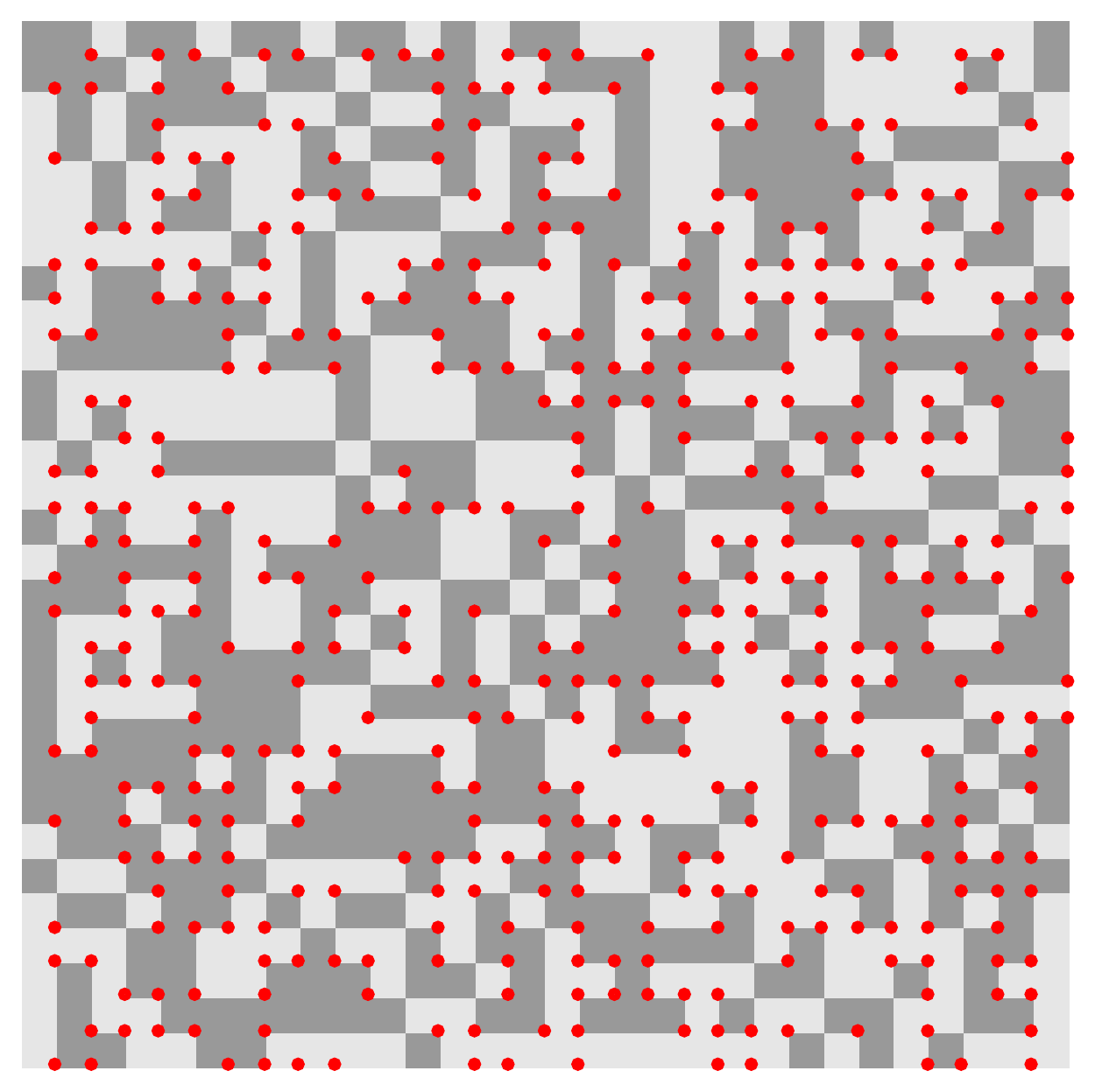}
\caption{Four states of the model with $N=N_F=30$. Dark squares are spin up and light squares are spin down. Red dots indicate the location of local defects where an odd number of the four adjacent spins is pointing up. Top to bottom, left to right: a ground state ($E=-N^3/16$, no defects), low lying states with 20 defects and then 68 defects (energies of order $-N^3$) and finally a high temperature state ($E\approx -N^2/8$, many defects). \label{fig:disorder}}
\end{center}
\end{figure}
This spin participates in $NN_F$ terms in the Hamiltonian, but in $N+N_F-1$ terms it appears raised 
to an even power, so the energy gap this predicts is $(N_F-1-\alpha+1/N)/8$. We confirmed this numerically for various small systems. 

To discuss ordering, it is useful to consider the operator
\be
X_{A,b}^{11}=S^{A,b}_z S^{A,b+1}_z S^{A+1,b}_z S^{A+1,b+1}_z,
\ee
which involves four neighboring spins (we have imposed periodic boundary conditions such that $S^{N+1,b}_z=S^{1,b}_z$ and $S^{A,N_F+1}_z= S^{A,1}_z$). This operator is a term in the Hamiltonian (\ref{eq:isingH}) and it is invariant under flipping spins along rows and columns. It is necessary to consider four spins in order to obtain an operator that is invariant under flipping the sign of both rows and columns independently. We will define a local defect at $(A+\half,b+\half)$ to be characterized by the fact that it has $X_{A,b}^{11}=-1$, which implies that one spin has the opposite sign of the other three spins. The ground states have $X_{A,b}^{11}=+1$ for all $A,b$. That is, the ground states do not have any local defects. The lowest excited states have four local defects, surrounding the flipped spin. Highly excited states are expected to have a lot of local defects. States with varying numbers of defects are illustrated in figure 
\ref{fig:disorder}.

The local defects are nice because they are easily visualized, but since the Hamiltonian is nonlocal, we could equally well have defined defects by the value of the generalized operator
\be
X_{A,b}^{M,n}=S^{A,b}_z S^{A,b+n}_z S^{A+M,b}_z S^{A+M,b+n}_z \,.
\ee
This corresponds to four spins on the corners of a rectangle of width $M$ and height $n$.
The statements that ground states have no defects and excited states have increasingly many defects again holds. Therefore, all of these defect operators are at our disposal to construct a quantity that distinguishes the large $Q$ saddle (with states that are `close' to the ground states) from the matrix integral saddle. The quantity that does the job nicely turns out to be 
\be\label{eq:square}
\hat{X}_{AM}=\frac{1}{N_F^2} \sum_{b,n} X_{A,b}^{M-A,n-b} = \left( \frac{1}{N_F} \sum_b S^{Ab}_z S^{Mb}_z \right)^2.
\ee
Here we may take any two rows labelled by $A$ and $M$.
The quantity $\hat{X}_{AM}$ has the virtue of admitting a simple expression in terms of the $Q$ matrix
\bea
\langle  \hat{X}_{PM} \rangle &=& \frac{c_N}{Z} \sum_{\sigma \in \{\pm 1\}^{N^2}} \int dQ \ e^{- \frac{N}{\beta} Q_{AB} Q_{AB} } \frac{4}{N_F^2} \frac{\partial^2}{(\partial Q_{PM})^2} \left( e^{ \frac{1}{2} \,  Q_{AB} \s^{Ac} \s^{Bc}} \right) \\
&=& \frac{c_N}{Z}  \int dQ \ \frac{4}{N_F^2} \frac{\partial^2}{(\partial Q_{PM})^2} \left( e^{- \frac{N}{\beta} Q_{AB} Q_{AB} } \right) \sum_{\sigma \in \{\pm 1\}^{N^2}} e^{ \frac{1}{2} \,  Q_{AB} \s^{Ac} \s^{Bc}}
\eea
\bea
&=& \frac{c_N}{Z}  \int dQ  \left( \frac{16}{\alpha^2 \beta^2}\  Q_{PM}^2 - \frac{8}{\a \beta} \frac{1}{N_F} \right) \ e^{- \frac{N}{\beta} Q_{AB} Q_{AB} + N_F \log (z(Q))} \\
& = &  \frac{16}{\alpha^2 \beta^2}\  \left\langle Q_{PM}^2 \right\rangle - \frac{8}{\a \beta} \frac{1}{N_F}  \,, \label{eq:Q2}
\eea
where $c_N=\frac{1}{2^{N/2}} \left(\frac{2 N}{\pi\beta}\right)^{N(N+1)/4}$. In the second line we used integration by parts. Note that the indices $P$ and $M$ are not summed over.

Using (\ref{eq:Q2}), it is now straightforward to evaluate $\langle  \hat{X}_{PM} \rangle$ in both saddles. In the matrix integral saddle, as we discussed above, the matrix elements of $Q$ are of order $\sqrt{\beta/N}$. In contrast, evaluated on the $Q$ matrices with large eigenvalues we found in (\ref{eq:largeQavg}) that the elements were of order $\a \beta$. Therefore we have that
\be\label{eq:niceX}
\left. \langle  \hat{X}_{PM} \rangle\right|_{\text{entrop.}} \sim \frac{1}{\a\b} \frac{1}{N_F} \to 0 \,, \qquad \qquad \left. \langle  \hat{X}_{PM} \rangle\right|_{\text{ener.}} \sim 1 \,.
\ee
That is, the expectation value of $\hat{X}$ cleanly distinguishes the two saddles in the large $N$ limit. Here -- following the discussion of section \ref{sec:meta} -- we have referred to the matrix model saddle as the `entropic' saddle while the large eigenvalue saddle is the `energetic' saddle.

$\hat X$ is not an order parameter in the sense that it is invariant under the symmetries of flipping the spins along rows and columns. Instead, it is better to think of $\hat{X}$ as the `square' of the order parameter (see (\ref{eq:square})), analogous to the spin-spin correlation function in the Ising model. In the usual local Ising model, `long range order' is characterized by a nonvanishing correlation between infinitely separated spins. In our nonlocal model, we see that the ordered phase can be characterized by a certain correlation between four spins at arbitrary separation. In the disordered phase, there is no correlation between these four spins, even when they are close together.  

The matrix integral phase is also characterized by an emergent $SO(N)$ symmetry. The quantity built from spins that admits a natural action of $SO(N)$ is $\sum_b S^{Ab}_z S^{Mb}_z$. Using manipulations like those of (\ref{eq:Q2}) above, correlators of this quantity can be expressed in terms of correlators of the $Q$ matrix. In the matrix integral phase, the effective action for $Q$ is $SO(N)$ invariant. This constraints correlators of $Q$ to be build from the $SO(N)$ invariant tensor $\delta_{AB}$. Thus, in a phase with an emergent $SO(N)$ symmetry, the correlators of $\sum_b S^{Ab}_z S^{Mb}_z$ are also constrained.

As well as an emergent $SO(N)$ symmetry, there can also be an emergent $SO(N_F)$ symmetry, that will now act on $\sum_A S^{Ab}_z S^{Ac}_z$. The entire discussion of the paper goes through if we exchange the role of $N$ and $N_F$ and therefore let $\alpha \to 1/\alpha$, which is now greater than one. Figure \ref{fig:Tcrit} can be extended to $\alpha > 1$, and so in particular we see that the range of temperatures over which $SO(N)$ and $SO(N_F)$ symmetries can emerge in a given system will not be the same in general. The two symmetries are not manifest simultaneously.

\section{Discussion}

Our system has a finite dimensional Hilbert space. The description in terms of continuous matrices appears only in the large $N$ expansion; there are no fundamental continuous degrees of freedom in the exact description. Perhaps interestingly, there are certain situations where it has been suggested that the underlying theory of a particular geometry, such as the static patch of de Sitter space \cite{Banks:2002wr, Witten:2001kn, Parikh:2004wh} or the AdS$_2 \times S^2$ near horizon of extremal black holes \cite{Verlinde:2004gt, Strominger:2003tm, Sen:2011cn}, may be described by a finite dimensional Hilbert space. If so, our system may serve as a greatly simplified toy model of such a situation. Moreover, both the static patch of de Sitter space and AdS$_2 \times S^2$, do not exist for indefinite temperatures and oftentimes are argued to be metastable. Curiously, and also perhaps interestingly, our system enjoys similar properties. Namely, the matrix integral phase only exists above a certain temperature and moreover is metastable over a significant range of temperatures.

To tighten the connection to interesting geometries, it will be important to study quantum versions of our system. Of interest are analogous matrix phases either at zero or non-zero temperature. In particular, an emergent matrix quantum mechanical phase at a quantum critical point might help elucidate the appearance of $SL(2,\mathbb{R})$ symmetries of the de Sitter static patch (and similarly for AdS$_2\times S^2$ geometries) studied in \cite{Anninos:2011af, Anninos:2013nra}. We also hope that a matrix quantum mechanics emergent from a spin system might be able to provide a microscopic realization of the connection between quantum entanglement and geometry suggested by the Ryu-Takayanagi formula for holographic entanglement entropy \cite{Ryu:2006bv, Swingle:2009bg}. From this point of view it might also be interesting to revisit the supersymmetric matrix models that in the planar limit map to a quantum spin chain \cite{Veneziano:2006}. Other matrix quantum mechanical systems with a finite Hilbert space are theories of purely fermionic matrices studied extensively in \cite{Semenoff:1996vm}.

We have focussed on the particular Hamiltonian (\ref{eq:isingH}). A natural objective for future work is to attain a more systematic understanding of when the $S_N$ and $\Z_2^N$ symmetries of spin systems are liable to become enhanced to an $SO(N)$ symmetry described by a single trace action. How much fine tuning is necessary in general? Relatedly, it will of course be important to find systems with more than one emergent matrix degree of freedom.

From a statistical physics perspective, one motivation to realize emergent matrix integral physics from spins was that the connectivity of the eigenvalue distribution is a type of topological order that can to lead to high order non-symmetry breaking phase transitions. In our model such a transition does not occur, because there is no stable disconnected eigenvalue saddle. However, the antiferromagnetic version of this model does exhibit a transition to a phase with a disconnected eigenvalue distribution, at a temperature above the onset of glassiness, although this point is not emphasized in \cite{parisi}.

\section*{Acknowledgements}
 
It is our pleasure to acknowledge helpful discussions with Frederik Denef, Jorge Santos, Steve Shenker and Tobias Tiecke.
This work is partially supported by a DOE early career award, by the Templeton foundation. D.A. acknowledges funding by the Martin A. and Helen Chooljian Founders� Circle and the National Science Foundation.

\appendix

\section{Quartic formulation of the matrix integral}
 \label{sec:quartic}
 
In this appendix we note that when $\alpha = 1$ the constrained matrix integral described by (\ref{eq:Zfinal}) and (\ref{eq:constraint}) can equivalently be formulated as a constrained quartic matrix integral. This explains the appearance of a known combinatorial expression \cite{math1, math2} for the on shell action in this case (\ref{eq:sexact}). To do this, all we have to do is adapt the approach
taken in \cite{parisi} to our `ferromagnetic' version of the matrix Ising model (\ref{eq:isingH}).

In \cite{parisi} it was shown that, in the high temperature phase, the partition function of the matrix Ising model is equivalent at large $N$ to the partition function of a matrix spherical spin model. The Hamiltonian for the matrix spherical spin model is identical to that of the matrix Ising model, but instead of Ising spins the degrees of freedom are real numbers $s^{Aa}$ subject to the global constraint $\sum_{A,a} |s^{Aa}|^2=N N_F$.

The equivalence between the matrix Ising model and the matrix spherical spin model in the high temperature phase also holds for our ferromagnetic matrix Ising model. The steps are those explained in \cite{parisi}. The case of $\alpha=1$, i.e. $N=N_F$, is especially simple. Because the spins now take values in the real numbers, we can directly diagonalize the matrix $s^{Aa}$ to express the spherical spin model partition function as an eigenvalue integral. In the large $N$ limit:
\beq
Z_{\textrm{sph}} \sim \int d \mu \int \Big( \prod_i d x_i \Big) e^{-N^2 S [\{ x_i, \mu \}]},
\eeq
where $\mu$ is introduced to impose the global constraint, the $x_i$ are the eigenvalues of the spin matrices and the action is given by
\beq \label{eq:sphaction}
S = \frac{1}{N} \sum_i \Big[- \frac{\beta}{16} x^4_i -\frac{\beta \mu}{32}x_i^2 -\frac{1}{2} \sum_{j \neq i} \log |x_i^2-x_j^2|  \Big] + \frac{\beta \mu}{32}.
\eeq
The first term comes directly from the 4-spin interaction (recall that $\lambda = -1/(16 N)$ in (\ref{eq:isingH})), the second and last term come from the constraint and the logarithmic term arises from the matrix transformation that diagonalizes the matrices over which we integrate. The action describes particles in a quartic potential with repulsive interactions through the logarithmic term. For $\alpha \neq1$ the main difference is that there will be an additional term $\log|x_i|$ in the above eigenvalue potential \cite{parisi}. This explains why these cases have more complicated expressions for the energy, they are not purely quartic. The external potential $- \frac{\beta}{16} x^4 -\frac{\beta \mu}{32}x^2$ has a local minimum for $\mu<0$ with a barrier height that is set by the temperature. The high temperature phase we are interested in thus becomes unstable below a critical temperature, where the eigenvalue repulsion pushes the eigenvalues over the barrier. This completely mimics the structure we found in the main text.

Introducing the symmetric eigenvalue distribution (normalized to unity)
\be
\rho(y) = \frac{1}{2N} \sum_i [ \delta(y - y_i) + \delta(y + y_i)] ,
\ee
the equations of motion derived from (\ref{eq:sphaction}) in the large $N$ limit become
\be
P\int \frac{dy \rho(y)}{x-y} = - \frac{\beta}{8} x^3 - \frac{\beta \mu}{32}  x .
\ee
The equation of motion for $\mu$ imposes the contraint
\beq\label{eq:muconstraint}
\int dy \rho(y) y^2 = 1.
\eeq
The following solution for the eigenvalue distribution is found
\beq
\rho(y) = - \beta \frac{\sqrt{b^2-y^2} \left(2 b^2+\mu + 4 y^2\right)}{32 \pi } \,, \qquad -b \leq y \leq b ,
\eeq
and vanishing support elsewhere. For the eigenvalue distribution to be nonnegative we need $\mu \leq -6 b^2$. Normalization and the constraint (\ref{eq:muconstraint}) from the equation of motion for $\mu$ result in the following equations for $b$ and $\mu$.
\bea
\beta b^2 \left(3 b^2+\mu \right)&=&-64 \,, \nonumber\\
\beta b^4 \left(4 b^2+\mu \right)&=&-256 \,. \nonumber
\eea
These equations yield real solutions for $b$ and $\mu$ provided that $\beta<\frac{16}{27}$. The nonnegative condition on the eigenvalue distribution is also obeyed in this temperature range. We thus reproduce the critical temperature $\beta_c=\frac{16}{27}$ found in (\ref{eq:criticalT}) in the main text.  Having obtained the eigenvalue distribution we can easily compute the energy as a function of temperature. We find perfect agreement with the results described in the main text and plotted in figure (\ref{fig:MC}).


\begin{thebibliography}{99}
 
\bibitem{'tHooft:1973jz} 
  G.~'t Hooft,
  ``A Planar Diagram Theory for Strong Interactions,''
  Nucl.\ Phys.\ B {\bf 72}, 461 (1974).
 
\bibitem{Klebanov:1991qa} 
  I.~R.~Klebanov,
  ``String theory in two-dimensions,''
  in {\it Trieste 1991, Proceedings, String theory and quantum gravity '91}, 30-101
  [hep-th/9108019].
 
\bibitem{Ginsparg:1993is} 
  P.~H.~Ginsparg and G.~W.~Moore,
  ``Lectures on 2-D gravity and 2-D string theory,''
  in {\it Boulder 1992, Proceedings, Recent directions in particle theory}, 277-469
  [hep-th/9304011].
  
\bibitem{Banks:1996vh} 
  T.~Banks, W.~Fischler, S.~H.~Shenker and L.~Susskind,
  ``M theory as a matrix model: A Conjecture,''
  Phys.\ Rev.\ D {\bf 55}, 5112 (1997)
  [hep-th/9610043].
  
\bibitem{Maldacena:1997re} 
  J.~M.~Maldacena,
  ``The Large N limit of superconformal field theories and supergravity,''
  Int.\ J.\ Theor.\ Phys.\  {\bf 38}, 1113 (1999)
  [Adv.\ Theor.\ Math.\ Phys.\  {\bf 2}, 231 (1998)]
  [hep-th/9711200].
 
\bibitem{Brezin:1977sv} 
  E.~Brezin, C.~Itzykson, G.~Parisi and J.~B.~Zuber,
  ``Planar Diagrams,''
  Commun.\ Math.\ Phys.\  {\bf 59}, 35 (1978).
  
\bibitem{Wen:2004ym} 
  X.~G.~Wen,
  {\it Quantum field theory of many-body systems: From the origin of sound to an origin of light and electrons}
  OUP (2004).
  
  \bibitem{parisi}
  L. F. Cugliandolo, J. Kurchan, G. Parisi, and F. Ritort,
  ``Matrix Models as Solvable Glass Models,''
  Phys. Rev. Lett. {\bf 74}, 1012 (1995).
  [cond-mat/9407086].
   
\bibitem{Denef:2011ee} 
  F.~Denef,
  ``TASI lectures on complex structures,''
 [arXiv:1104.0254 [hep-th]].
  
\bibitem{Masuku:2014wxa} 
  M.~Masuku, M.~Mulokwe and J.~P.~Rodrigues,
  ``Large N Matrix Hyperspheres and the Gauge-Gravity Correspondence,''
  [arXiv:1411.5786 [hep-th]].
  
  \bibitem{math1}
  D.~Bessis, C.~Itzykson and J.~B.~Zuber,
  ``Quantum field theory techniques in graphical enumeration,''
   Adv. Appl. Math. {\bf 1}, 109 (1980).
  
\bibitem{math2}
  A.~Zvonkin,
  ``Matrix integrals and map enumeration an accessible introduction,''
   Math. Comput. Modelling {\bf 26}, 281 (1997).

\bibitem{MCdos}
F.~Wang, and D. P.~Landau,
`` Efficient, Multiple-Range Random Walk Algorithm to Calculate the Density of States,"
Phys. Rev. Lett. {\bf 86}, 2050 (2001).

   
\bibitem{Banks:2002wr} 
  T.~Banks, W.~Fischler and S.~Paban,
  ``Recurrent nightmares? Measurement theory in de Sitter space,''
  JHEP {\bf 0212}, 062 (2002)
  [hep-th/0210160].

\bibitem{Witten:2001kn} 
  E.~Witten,
  ``Quantum gravity in de Sitter space,''
  [hep-th/0106109].

\bibitem{Parikh:2004wh} 
  M.~K.~Parikh and E.~P.~Verlinde,
  ``De Sitter holography with a finite number of states,''
  JHEP {\bf 0501}, 054 (2005)
  [hep-th/0410227].

\bibitem{Verlinde:2004gt} 
  H.~L.~Verlinde,
  ``Superstrings on AdS(2) and superconformal matrix quantum mechanics,''
  [hep-th/0403024].

\bibitem{Strominger:2003tm} 
  A.~Strominger,
  ``A Matrix model for AdS(2),''
  JHEP {\bf 0403}, 066 (2004)
  [hep-th/0312194].

\bibitem{Sen:2011cn} 
  A.~Sen,
  ``State Operator Correspondence and Entanglement in $AdS_2/CFT_1$,''
  Entropy {\bf 13}, 1305 (2011)
  [arXiv:1101.4254 [hep-th]].

\bibitem{Anninos:2011af} 
  D.~Anninos, S.~A.~Hartnoll and D.~M.~Hofman,
  ``Static Patch Solipsism: Conformal Symmetry of the de Sitter Worldline,''
  Class.\ Quant.\ Grav.\  {\bf 29}, 075002 (2012)
  [arXiv:1109.4942 [hep-th]].

\bibitem{Anninos:2013nra} 
  D.~Anninos, T.~Anous, P.~de Lange and G.~Konstantinidis,
  ``Conformal Quivers and Melting Molecules,''
  [arXiv:1310.7929 [hep-th]].
  
\bibitem{Ryu:2006bv} 
  S.~Ryu and T.~Takayanagi,
  ``Holographic derivation of entanglement entropy from AdS/CFT,''
  Phys.\ Rev.\ Lett.\  {\bf 96}, 181602 (2006)
  [hep-th/0603001].
  
\bibitem{Swingle:2009bg} 
  B.~Swingle,
  ``Entanglement Renormalization and Holography,''
  Phys.\ Rev.\ D {\bf 86}, 065007 (2012)
  [arXiv:0905.1317 [cond-mat.str-el]].

\bibitem{Veneziano:2006}
G. Veneziano and J. Wosiek,
``A supersymmetric matrix model: III. Hidden SUSY in statistical systems,"
JHEP {\bf 11}, 030 (2006)
[hep-th/0609210]
  
\bibitem{Semenoff:1996vm} 
  G.~W.~Semenoff and R.~J.~Szabo,
  ``Fermionic matrix models,''
  Int.\ J.\ Mod.\ Phys.\ A {\bf 12}, 2135 (1997)
  [hep-th/9605140].




 
 \end{thebibliography}
\end{document}